# Ferromagnet in a continuously tuneable random field

D.M. Silevitch[1], D. Bitko[2], J. Brooke[3], S. Ghosh[4], G. Aeppli[5], T. F. Rosenbaum[1]

[1]*The James Franck Institute and Department of Physics, The University of Chicago, Chicago, IL 60637*

[2]*MPI Research Inc., East Brunswick, NJ 08816*

[3]*Lincoln Laboratory, Massachusetts Institute of Technology, Lexington, MA 02420*

[4]*School of Natural Sciences, University of California Merced, Merced, CA 95344*

[5]*London Centre for Nanotechnology and Department of Physics and Astronomy, UCL, London, WC1E 6BT, UK*

**Most physical and biological systems are disordered, even while the majority of theoretical models treat disorder as a weak perturbation. One particularly simple system is a ferromagnet approaching its Curie temperature, $T_C$, where all of the spins associated with partially-filled atomic shells acquire parallel orientation. With the addition of disorder via chemical substitution, the Curie point is suppressed, but no qualitatively new phenomena appear in bulk measurements as long as the disorder is truly random on the atomic scale and not so large as to eliminate ferromagnetism entirely.[1] Here we report the discovery that a simply measured magnetic response is singular above the Curie temperature of a model, disordered magnet, and that the associated singularity grows to an anomalous divergence at $T_C$. The origin of the singular response is the random internal field induced by an external magnetic field transverse to the favoured direction for magnetization.[2-4] The fact that ferromagnets can be studied easily and with high-precision using bulk susceptibility and a large variety of imaging tools will not only advance fundamental studies of the random field problem, but also suggests a mechanism for tuning the strength of domain wall pinning, the key to applications.**

Ferromagnets display many interesting effects when exposed to external magnetic fields $H$. Most important for applications from motors to disk drive read heads is the switching and hysteresis that occurs when the magnetic field changes sign. Magnetic domain-wall pinning makes the magnetization $M$ evolve continuously around an $H$-dependent loop rather than simply jumping at $H=0$, as expected without pinning. For fundamental statistical mechanics, equally important are the temperature dependence of $M$ above $T_C$ for an infinitesimal field $dH$ and the $H$-dependence of $M$ at $T=T_C$. For many ferromagnets, these are both singular, in that they can be described by power laws in the reduced temperature $T-T_C$ and the field $H$ respectively. Thus, the magnetic susceptibility $\chi=dM/dH$ diverges like $(T-T)^{-\gamma}$ for small $H$ and like $H^{1/\delta-1}$ for $T=T_C$. In the last century, these divergences drove vast experimental and theoretical efforts, leading to the formulation of the concept of scale invariance at second order phase transitions.[5]

Real-world magnets depart from the ideal system treated by statistical mechanics, typically being riddled with structural and chemical defects that are often deliberately introduced to engineer specific properties. Recognizing this, theorists have produced many models incorporating simplified representations of the disorder seen in realistic magnets. One useful generic model is the "random-field model", where in addition to a uniform external field, there is a field which varies randomly from site to site with zero mean. Site-random fields are difficult to tune experimentally, and the random field model seemed quite abstract and not amenable to quantitative test until Aharony and Fishman[6] proposed that site-random antiferromagnets in tuneable, uniform external magnetic fields should behave like ferromagnets in tuneable random fields (Fig. 1A). This insight set off a flurry of activity about 25 years ago[7-12] because insulating antiferromagnets can be characterized optically and via magnetic neutron scattering. Important results were that for small random fields, high dimensional (*d*) magnets retain their order, and that under these circumstances, the critical exponent γ changes from its clean limit value of 1.25[13] to 1.58 for *d*=3.[14]

Recently, various authors[2-4] have emphasized that there are also ferromagnets for which an external dc field generates an effective internal random field (Fig. 1A). This provides a new venue for investigating the random field problem, with the advantages of a more obvious relation to technical ferromagnetism, the availability of direct bulk probes such as magnetization and susceptibility, and the ability to measure directly the behaviour of $\chi(H)$ for $T \geq T_C$. It is important to note that this direct approach is inapplicable to antiferromagnets because of the inability to apply a staggered (i.e. regularly alternating in space) field.

The starting point is the Ising model in a transverse field for a disordered magnet, with Hamiltonian $H = -\sum_{\langle ij \rangle} J_{ij} S_i^z S_j^z - \Gamma \sum_i S_i^x$ where $S_i$ is the spin operator at site *i* and $J_{ij}$ represents the interaction between spins at sites *i* and *j*. The transverse field $\Gamma$ leads to quantum mechanical mixing of the Ising (**z**) up and down spin states at each site, with the outcome that eventually there is a well-known quantum critical point at *T*=0 for $\Gamma \sim J_{ij}$ beyond which magnetic order along **z** ceases to survive (see Fig. 1B). $\Gamma$ also generates a non-vanishing expectation value $\langle S_i^x \rangle$ at each site, which will exert a field on the Ising components $S_j^z$ of the spin operator if an interaction with off-diagonal (in spin space) terms is added to the Hamiltonian and these terms do not cancel by symmetry. An example of such an interaction is the magnetic dipolar coupling for a ferromagnet simply disordered by site dilution, illustrated in Fig. 1A.

For our experiments on the new route to the random field Ising model, we use $LiHo_xY_{1-x}F_4$, a transparent insulator. Reference 15 contains a recent description of the underlying Hamiltonian, including crystal fields, a laboratory magnetic field $H$, and nuclear hyperfine interactions, and their detailed consequences for the magnetic dynamics. The $Ho^{3+}$ ions carry large magnetic moments, strongly aligned along the **c**-axis of the tetragonal material and coupled to each other via the dipolar interaction. The outcome is that pure $LiHoF_4$ is an effective spin-1/2 Ising ferromagnet with a Curie temperature $T_C$ = 1.53 K.[16] Fields $H_t$ transverse to **c** convert the pure compound into a realization of the transverse field Ising model, with a continuous phase boundary between the $H_t$=0 classical critical point and a $T$=0 quantum critical point at $H_t$=50 kOe.[17] For $H_t$<20 kOe, $\Gamma$ is quadratic in $H_t$.[18] Mean field theory, taking into account both the electronic and nuclear spin degrees of freedom, quantitatively describes the static and dynamical properties in the $H_t$-$T$ plane, including the classical ($H_t \to 0$) and quantum ($T \to 0$) limits.[15,17,19] When non-magnetic yttrium partially replaces magnetic holmium, the classical disordered ferromagnet orders at a depressed $T_C(x) = xT_C(x=1)$ down to x ~ 0.2, where the combination of disorder and frustration arising from the spatial anisotropy of the dipole interaction gives rise to a spin glass ground state.[1]

While the measured Curie point has precisely the value anticipated in the simplest molecular field theory, the quantum critical point occurs at greatly suppressed values, as can be seen in the phase diagrams (two of which have been published previously[17,20]) of Fig. 1B. Indeed, it was the discrepancy between simple theory and these data that motivated the two recent papers[2,3] that pointed out the generation of internal random fields in disordered dipole-coupled Ising magnets. In this work, we focus on the region near and above the ferromagnetic $T_C(x)$ to explore the random-field induced change in the classical, "high-temperature" critical behaviour. By contrast, our previous work on the low-temperature quantum behaviour of the same materials focused on different phenomena, including the quantum glass[18], entanglement (caused by the same off-diagonal terms in the dipole interaction as are responsible for the random fields discussed here)[4] and antiglass behaviour[21], decoherence[15], tuneable domain wall tunneling[22] and the associated concept of quantum annealing.[20]

Fig. 2 shows the evolution of the raw $\chi'(\Gamma)$ data (see *Experimental Methods*) with decreasing $T$ for the most disordered of our ferromagnetic samples, namely the x=0.44 crystal. At high $T$, the susceptibility is sharply peaked, and its divergence at the Curie point is only cut off by demagnetizing effects due to sample geometry. The extraordinary sharpness of the susceptibility peaks again confirms the ideal nature of $LiHo_xY_{1-x}F_4$ as a model system for studying the combined effects of quantum mechanics and quenched disorder. For $T < 0.300$ K, however, the peak becomes significantly broader and rounder, and the peak susceptibility never reaches the demagnetization limit. By contrast, comparable measurements on pure $LiHoF_4$ reach the demagnetization limit for temperatures as low as 0.025 K.[17,23] As we have shown previously,[20,22] the system becomes a randomly pinned ferromagnetic domain state and in many respects behaves like a glass, unable to reach equilibrium over the measurement time, yet retains a net moment as indicated by local Hall-probe magnetometry. These results are in excellent qualitative agreement with the concept that with growing transverse field, the induced random field becomes large enough to prevent the achievement of an equilibrium ferromagnetic state, in exact analogy with what had previously been found for the random Ising antiferromagnets in external fields.

We now focus on the critical behaviour for x=1 and x=0.44 on the approach to the classical critical point $T_C(\Gamma=0)$. Fig. 3 shows the susceptibility for the two concentrations near $T_C(\Gamma=0)$ as a function of $T$ at $\Gamma=0$, as well as $\Gamma$ at $T=T_C$. Both the ordered (x=1) and disordered (x=0.44) crystals show power-law behaviour $\chi' \sim T^{-\gamma}$ over multiple decades as a function of temperature. In both cases, the thermal critical exponent was measured to be $\gamma=1.00\pm0.04$, in accordance with the mean-field prediction.[24-26] As a function of $\Gamma$, the ordered system shows an apparent critical exponent of $1/\delta_\Gamma-1=-0.99\pm0.0005$. The low-$\Gamma$ asymptote is to be contrasted with $\chi' \sim \Gamma^{-2}$, given by mean field theory for the transverse field Ising model when the hyperfine interactions are either negligible or much larger than $\Gamma$. For $Ho^{3+}$, the hyperfine coupling $A=39$ mK[15], and because the nuclear and electronic spins are large (7/2 and 8 respectively), the bandwidth of nuclear excitations is therefore also large. This accounts for the bending of the susceptibility towards very shallow behaviour as $\Gamma$ goes beyond A. At the same time, for the smallest $\Gamma$, we will be dealing with both absolute uncertainties and fluctuations in the temperature, meaning that the expected $1/\Gamma^2$ singularity will be smoothed out to yield the less singular, but still strongly divergent $1/\Gamma$ form that we actually observe.

The disordered system demonstrates qualitatively different behaviour (Fig. 3b). Nearest the classical critical point, we observe power-law behaviour with an exponent of $1/\delta_\Gamma-1=-0.57\pm0.03$, with a crossover at larger $\Gamma$ to $1/\delta_\Gamma-1=-1$. We emphasize that these are entirely intrinsic results, and while the hyperfine interactions clearly remain a factor, thermal broadening is not an issue here as it is for very small $\Gamma$ for the pure compound, where the phase boundary rises very steeply. There is an even more remarkable result at temperatures above $T_C$. Fig. 4 reveals non-analytic behaviour of $\chi'(H_t)$ in that there is no rounding as $H_t \rightarrow 0$ (note that $\chi'$, measured along **c**, is an even function of $H_t$, peaked at $H_t=0$); a prior study on a more dilute and non-ferromagnetic (x=0.17) concentration observed a hint of a cusp at $H_t=0$ rather than a rounded maximum at $H_t=0$.[17] Mathematically, this means that we must add an anomalous linear term $a|H_t|$ to the expansion $\chi(H_t)=\chi(T,H_t=0)+ bH_t^2+\ldots$. That the cusp at $H_t=0$ is a maximum rather than a minimum implies that the coefficient $a$ of the new term must be negative, as is that of the quadratic term. There is consequently an added suppression of the Ising correlations due to the disorder, and it is precisely given by a term of first order in the amplitude $|H_t|$ of the random field generated by $H_t$.

The $T>T_C$ singularity $\chi'\sim\chi'_0-a|H_t|$, has the same leading order (in $|H_t|$) behaviour as the Curie-Weiss-like expression $\chi'\sim 1/(c_1+c_2|H_t|)$, which itself is consistent with the measured exponent $\gamma\sim 1/2$ found at $T_C$ (when $c_1=0$) because $\Gamma\sim H_t^2$. We therefore can encapsulate both the $T>T_C$ singularity and the anomalous exponent at $T_C$ in the following modified Curie law for the susceptibility per Ho atom:

$$\chi' = \frac{C}{\alpha'\mu_B|H_t|+(T-T_C)+\gamma'\Gamma}, \qquad (1)$$

where the term proportional to $\Gamma$ takes account of the quantum fluctuations in the disordered system as well as higher-order random field effects, and C is the Curie constant for an isolated $Ho^{3+}$ ion in the fluoride. The parameters $\alpha'_{0.44}=0.157\pm 0.001$ and $\gamma'_{0.44}=0.321\pm 0.001$ were determined by fitting $\chi'$ ($\Gamma$) at $T=T_C$ to Eq. 1, as shown in Fig 3B.

As shown in the inset of Fig 4, the measured susceptibilities for temperatures above and below $T_C$ collapse onto a single universal curve following Eq. 1, using the values of $\alpha'$ and $\gamma'$ determined from Fig 3B. Indeed, to within experimental accuracy, Eq. 1 is an exact description of the susceptibility at the classical critical point, with $T_C=0.669$ K and $H_t=0$ for x=0.44. The loci in $\Gamma$-$T$ space where Eq. 1 diverges are also consistent with the direct measurement of the phase boundary shown in Fig. 1, which is linear in $\Gamma$ from 0.2 K to 0.625 K, with a crossover to linear in $|H_t|$ approaching the classical critical point.

Our experiments are the first to approach the critical behaviour of the random field Ising problem in a ferromagnet where the control parameter is the random field amplitude itself. Over many decades of reduced temperature and random field amplitude, a generalized Curie-Weiss form, Eq. 1, describes the data. This is perhaps not surprising given that Curie-Weiss descriptions inevitably arise in molecular field theories applicable when interactions are long ranged, like the dominant dipolar interaction in Li(Ho,Y)F$_4$. What is unexpected, however, is that the leading order contribution of the random field term is $|H_t|$ rather than $H_t^2$; any perturbative approaches that we have used to determine the random field effects on $\chi$ yield only even order terms in $H_t$. The only means for generating the linear term are for rare, large amplitude random fields to become so important that they destroy the viability of perturbation expansions in the product of the off-diagonal interaction term and $H_t$. Our detailed results for $\chi$, therefore, are a very dramatic – and indeed unprecedented – manifestation of "Griffiths singularities".[27] A particularly simple outcome of the Griffiths singularities is the shape of the phase diagram $T_C(\Gamma)$ terminating at the quantum critical point ($T=0$ and $\Gamma=\Gamma_C$), which does not agree with mean field theory for $\Gamma>0$. The divergence of $\chi'$ in Eq.(1) for our x=0.44 samples occurs at a line of Curie points where $T_C(H_t) - T_C(H_t = 0) = -\alpha' \mu_B |H_t| - \gamma' \Gamma = -0.16 \mu_B |H_t| - 0.32 \Gamma$, implying a low field asymptote of $T_C(H_t) - T_C(H_t = 0) \sim |H_t|$. For the less dilute sample (x= 0.65), it appears that $T_C(H_t) - T_C(H_t = 0) \sim H_t^2$, in agreement with perturbation theory. The numerical results of Ref. 2, while covering a much coarser grid in the $H_t$-$T$ plane than our experiments, are consistent with the appearance of the linear term with larger dilution.

We have implemented a new method to execute quantitative studies of the long-established random field problem, and have immediately made surprising discoveries using conventional magnetic susceptometry. This opens the door to investigation of the random field problem by other techniques, many of which, such as magnetic noise spectroscopy, magneto-optics and magnetic force microscopy, are much better matched to ferromagnets than antiferromagnets. This will result in the ability to examine both the statics and dynamics of the random-field problem, and perform detailed checks of long-standing theoretical predictions.

**Methods**

We performed ac susceptibility measurements on single crystals of LiHo$_x$Y$_{1-x}$F$_4$ with x=1.0, 0.65, and 0.44. The (5x5x10) mm$^3$ single-crystal samples with long axis along **c** were mounted on the cold finger of a helium dilution refrigerator placed inside a 8T superconducting magnet with field transverse to the Ising axis. AC excitation fields of 20-50 mOe at 10 Hz (chosen to ensure linear response ) were applied along the Ising axis with the magnetic response recorded using an inductive pickup coil in a gradiometer configuration.

REFERENCES

[1] Reich, D.H. *et al.*, Dipolar magnets and glasses: Neutron-scattering, dynamical, and calorimetric studies of randomly distributed Ising spins. *Phys. Rev. B* **42**, 4631-4644 (1990).

[2] Tabei, S.M.A., Gingras, M.J.P., Kao, Y.-J., Stasiak, P. & Fortin, J.-Y., Induced Random Fields in the $LiHo_xY_{1-x}F_4$ Quantum Ising Magnet in a Transverse Magnetic Field. *Phys. Rev. Lett.* **97**, 237203 (2006).

[3] Schechter, M., $LiHo_xY_{1-x}F_4$ as a random field Ising ferromagnet, cond-mat/0611063 (2006).

[4] Ghosh, S., Rosenbaum, T. F., Aeppli, G.& Coppersmith, S. N., Entangled quantum state of magnetic dipoles. *Nature* **425**, 48-51 (2003).

[5] Ma, S.K., *Modern Theory of Critical Phenomena* (Addison-Wesley, Reading, Mass, 1976).

[6] Fishman, S. & Aharony, A., Random Field Effects In Disordered Anisotropic Anti-Ferromagnets. *J. Phys. C* **12**, L729-733 (1979).

[7] Gofman, M. *et al.*, Critical behavior of the random-field Ising model. *Phys. Rev. B* **53**, 6362-6384 (1996).

[8] Yoshizawa, H. *et al.*, Random-Field Effects in Two- and Three-Dimensional Ising Antiferromagnets. *Phys. Rev. Lett.* **48**, 438-441 (1982).

[9] Ferreira, I.B., King, A.R., Jaccarino, V., Cardy, J.L. & Guggenheim, H.J., Random-field-induced destruction of the phase transition of a diluted two-dimensional Ising antiferromagnet: $Rb_2Co_{0.85}Mg_{0.15}F_4$. *Phys. Rev. B* **28**, 5192-5198 (1983).

[10]Belanger, D.P., King, A.R., Jaccarino, V. & Cardy, J.L., Random-field critical behavior of a d=3 Ising system. *Phys. Rev. B* **28**, 2522-2526 (1983).

[11]Birgeneau, R.J. *et al.*, Critical behavior of a site-diluted three-dimensional Ising magnet. *Phys. Rev. B* **27**, 6747-6753 (1983).

[12]Belanger, D.P. & Young, A.P, The Random Field Ising-Model. *J Magn. Magn. Mater*. **100**, 272-291 (1991).

[13]Belanger, D.P. & Yoshizawa, H., Neutron scattering and the critical behavior of the three-dimensional Ising magnet $FeF_2$. *Phys. Rev. B* **35**, 4823-4830 (1987).

[14]Slanic, Z, Belanger D.P. & Fernandez-Baca, J.A Equilibrium Random-Field Ising Critical Scattering in the Antiferromagnet $Fe_{0.93}Zn_{0.07}F_2$. *Phys. Rev. Lett*. **82**, 426-429 (1999).

[15]Ronnow, H.M. *et al.*, Quantum Phase Transition of a Magnet in a Spin Bath. *Science* **308**, 389-392 (2005).

[16]Hansen, P.E., Johansson, T. & Nevald, R., Magnetic properties of lithium rare-earth fluorides: Ferromagnetism in $LiErF_4$ and $LiHoF_4$ and crystal-field parameters at the rare-earth and Li sites. *Phys. Rev. B* **12**, 5315-5324 (1975).

[17]Bitko, D., Rosenbaum, T.F. & Aeppli, G*.,* Quantum Critical Behavior for a Model Magnet. *Phys. Rev. Lett*. **77**, 940-943 (1996).

[18]Wu, W., Ellman, B., Rosenbaum, T.F., Aeppli, G. & Reich, D.H., From classical to quantum glass. *Phys. Rev. Lett*. **67**, 2076-2079 (1991).

[19]Chakraborty, P.B., Henelius, P. Kjonsberg, H., Sandvik, A.W. & Girvin, S.M., Theory of the magnetic phase diagram of $LiHoF_4$. *Phys. Rev. B* **70**, 144411 (2004).

[20]Brooke, J., Bitko, D., Rosenbaum, T.F. & Aeppli, G., Quantum Annealing of a Disordered Magnet. *Science* **284**, 779-781 (1999).

[21]Ghosh, S., Parthasarathy, R., Rosenbaum, T.F. & Aeppli, G., Coherent Spin Oscillations in a Disordered Magnet. *Science* **296**, 2195-2198 (2002).

[22]Brooke, J., Rosenbaum, T.F. & Aeppli, G., Tunable quantum tunnelling of magnetic domain walls. *Nature* **413**, 610-613 (2001).

[23]Bitko, D., *Order and Disorder in a Model Quantum Magnet* (PhD Thesis, The University of Chicago, Chicago, Illinois, 1997).

[24]Aharony, A. & Halperin, B.I, Exact Relations among Amplitudes at Critical Points of Marginal Dimensionality. *Phys. Rev. Lett.* **35**, 1308-1310 (1975)

[25]Brezin, E. & Zinn-Justin, J., Critical behavior of uniaxial systems with strong dipolar interactions. *Phys. Rev. B* **13**, 251-254 (1976)

[26]Ahlers, G., Kornblit, A. & Guggenheim, H.J., Logarithmic Corrections to the Landau Specific Heat near the Curie Temperature of the Dipolar Ising Ferromagnet $LiTbF_4$. *Phys. Rev. Lett.* **34**, 1227-1230 (1975).

[27]Griffiths R.B., Nonanalytic Behavior Above the Critical Point in a Random Ising Ferromagnet. *Phys. Rev. Lett* **23**, 17-19 (1969).

Reprints and permissions information is available at npg.nature.com/reprintsandpermissions

**Acknowledgements** The work at the University of Chicago was supported by the US Department of Energy and the MRSEC program of the National Science Foundation, while work in London was supported via the UK Engineering and Physical Sciences Research Council and a Wolfson-Royal Society Research Merit Award.

**Competing Interest Statement** The authors declare that they have no competing financial interests.

**Correspondence** and requests for materials should be addressed to tfr@uchicago.edu.

FIGURE CAPTIONS

**Fig. 1** Random fields in a diluted, dipolar-coupled ferromagnet drive the system away from mean-field behavior. **a** Schematic showing two realizations of the random-field model. Top row: Aharony-Fishman method[2], utilizing a site-diluted antiferromagnet. Bottom row: Dilute dipole-coupled Ising ferromagnet[11-13], where the off-diagonal terms of the dipole interaction act to enable the random fields. In both cases, the undiluted system (left column) experiences no random-field effects due to overall symmetry; breaking the symmetry by dilution with non-magnetic sites (right column) introduces a net random field. **b** Normalized ferromagnetic-paramagnetic phase diagram for $LiHo_xY_{1-x}F_4$. From top to bottom, x=1.0, 0.65, and 0.44, with $T_C$ = 1.53, 1.02 and 0.669 K, respectively. For x=1, the solid line is derived from mean field theory[15] including nuclear and electronic degrees of freedom, while for the other compositions, the curves are guides for the eye. The normalization constants $T_C$ and $\Gamma_C$ are the values $xT_C(x=1)$ and $\Gamma_C=x\Gamma_C(x=1)$ predicted from mean-field theory at $\Gamma$=0 and T=0 respectively. The upturn for T < 0.4 K reflects the influence of the hyperfine interaction where the coupling between electronic (J) and nuclear (I) spins creates a larger effective moment (I+J) **Inset**: Phase boundary for x=0.44 in absolute units. Solid line is shows the phase boundary predicted by mean field theory, using the parameters derived from the model for x=1, with the average spin-spin coupling strength J scaled by the concentration. Dashed line shows the phase boundary derived from the critical divergences (see text for details). Dashed lines with arrows are where the critical curves plotted in Fig 3B were measured.

**Fig. 2** Real part of the magnetic susceptibility $\chi'$ of LiHo$_{0.44}$Y$_{0.56}$F$_4$ measured vs. transverse field $\Gamma$ at a series of temperatures (T = 0.650, 0.625, 0.600, 0.500, 0.400, 0.300, 0.200, 0.150, 0.100, 0.080, 0.060, and 0.025 K). Below T = 0.300 K the cusp in $\chi'$ becomes rounded at the Curie point and the response no longer reaches the demagnetization limit as glassy effects set in.

**Fig. 3** Critical behaviour of LiHo$_x$Y$_{1-x}$F$_4$ in the paramagnetic regime as a function of T at $\Gamma$=0 and $\Gamma$ at T=T$_C$. Lines are power law fits described in text. **a** x=1.00 (T$_C$=1.53 K). **b** x=0.44 (T$_C$=0.669 K). $\chi$(T) shows mean-field behaviour, whereas $\chi(\Gamma)$ exhibits a suppressed critical exponent indicative of the effects of the random fields, fitted to Eq. 1 in the text. For both concentrations, absolute error in T$_C$ is ±0.2 mK with a thermal stability of ±0.02 mK.

**Fig. 4** Singular behaviour of the susceptibility of LiHo$_{0.44}$Y$_{0.56}$F$_4$ at T=0.673 K (0.004 K above T$_C$). **Inset**: Inverse susceptibility at a series of temperatures above, at and below T$_C$, collapsed onto a universal, singular curve using the fitting parameters derived in Fig. 3.

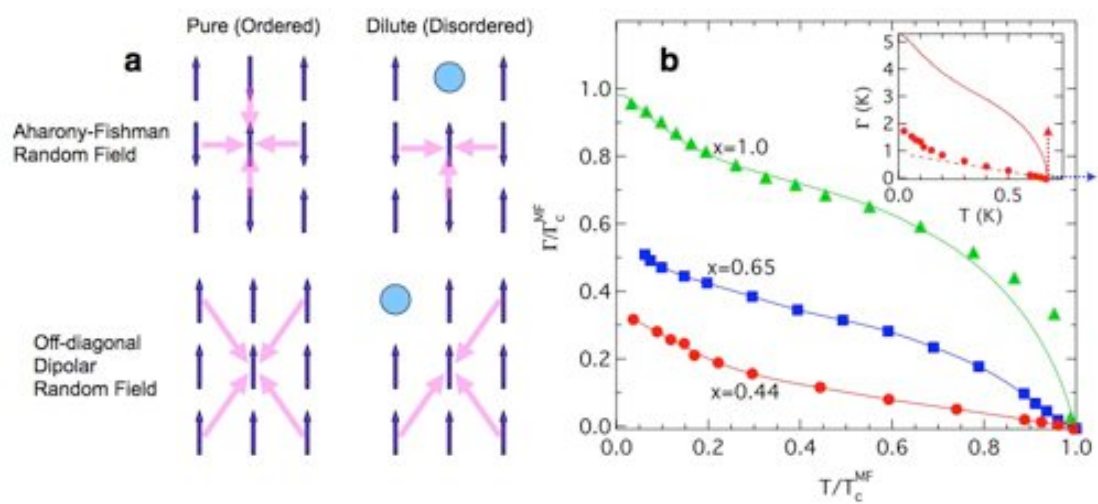

Figure 1

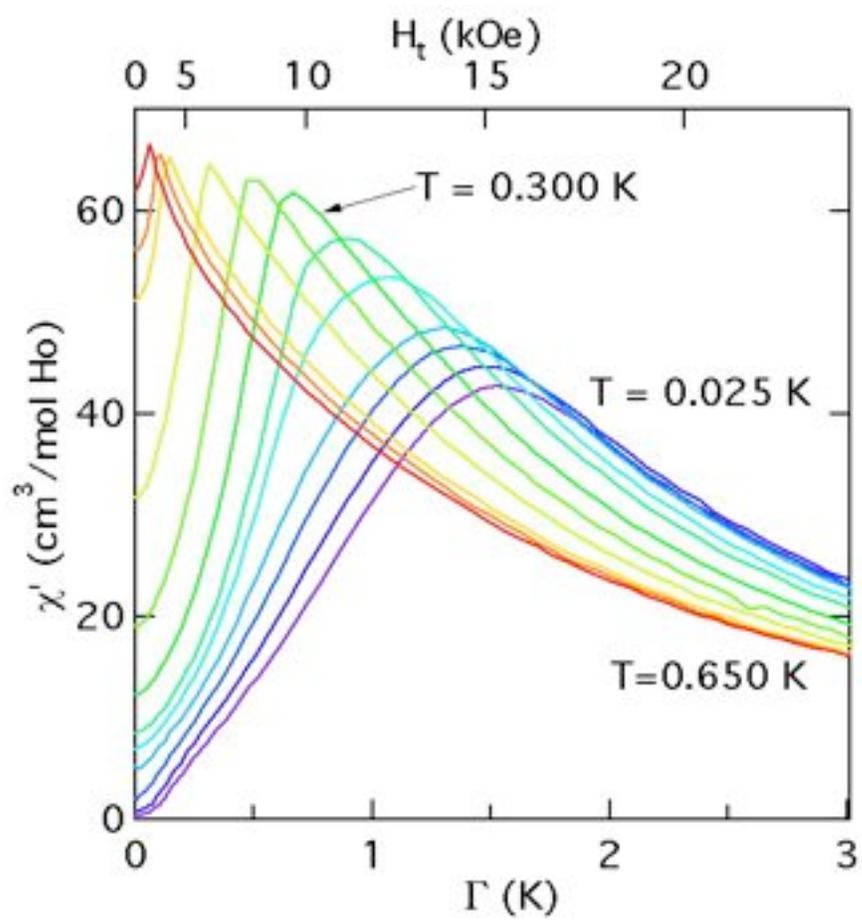

Figure 2

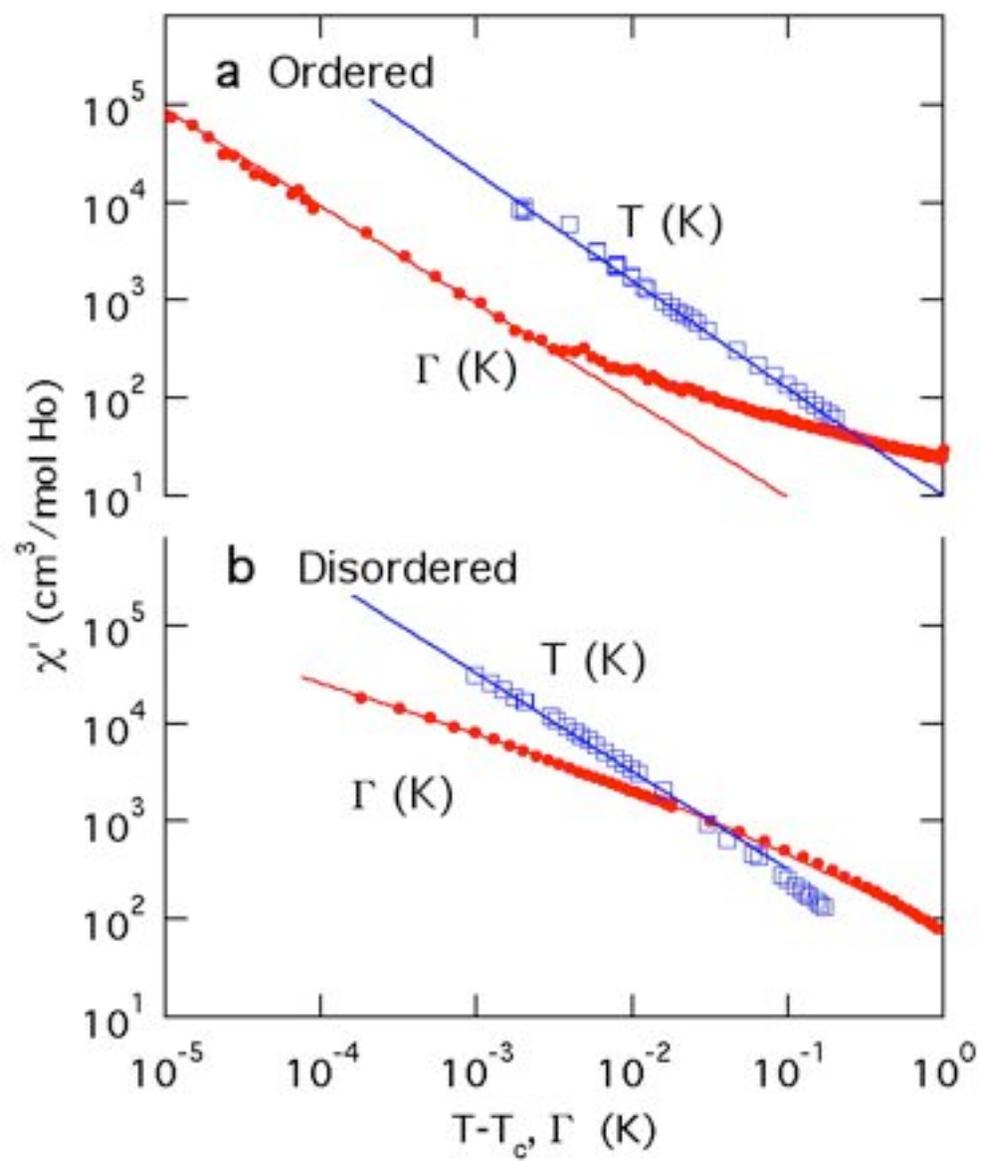

Figure 3

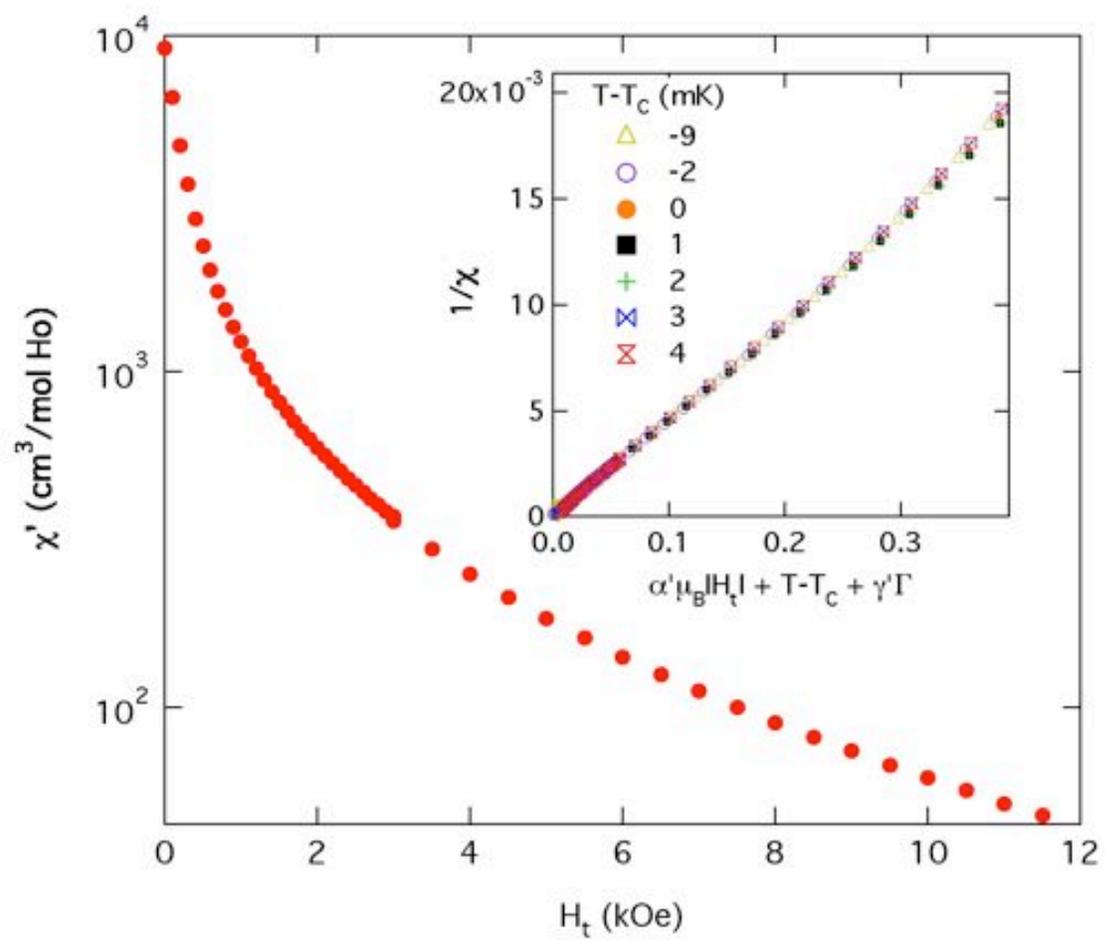

Figure 4